\documentclass[usenatbib]{mnras}

\usepackage[T1]{fontenc}
\usepackage{ae,aecompl}

\usepackage{graphicx}	
\usepackage{amssymb}
\usepackage[fleqn]{amsmath}

\newcommand{\ujypbm}{$\mu$Jy\,beam$^{-1}$}
\newcommand{\rad}{\texttt{RADMC-3D }}

\begin{document}

\title[ALMA Observations of HD 38206]{Resolving the outer ring of HD 38206 using ALMA and constraining limits on planets in the system
}
\author[M. Booth et al.]{Mark Booth$^{1}$\thanks{E-mail: markbooth@cantab.net}, Michael Schulz$^{1}$, Alexander V. Krivov$^{1}$, Sebasti\'an Marino$^{2,3}$, \newauthor Tim D. Pearce$^1$ and Ralf Launhardt$^2$\\
$^{1}$ Astrophysikalisches Institut und Universit\"atssternwarte, Friedrich-Schiller-Universit\"at Jena, Schillerg\"a\ss{}chen 2-3, 07745 Jena, \\Germany \\
$^2$ Max Planck Institute for Astronomy, K\"onigstuhl 17, D-69117 Heidelberg, Germany \\
$^3$ Institute of Astronomy, University of Cambridge, Madingley Road, Cambridge CB3 0HA, UK \\
}

\date{Accepted 2020 October 23. Received 2020 October 21; in original form 2020 September 1}

\maketitle

\begin{abstract}
HD~38206 is an A0V star in the Columba association, hosting a debris disc first discovered by \emph{IRAS}. Further observations by \emph{Spitzer} and \emph{Herschel} showed that the disc has two components, likely analogous to the asteroid and Kuiper belts of the Solar System. The young age of this star makes it a prime target for direct imaging planet searches. Possible planets in the system can be constrained using the debris disc. Here we present the first ALMA observations of the system's Kuiper belt and fit them using a forward modelling MCMC approach. We detect an extended disc of dust peaking at around 180~au with a width of 140~au. The disc is close to edge on and shows tentative signs of an asymmetry best fit by an eccentricity of $0.25^{+0.10}_{-0.09}$. We use the fitted parameters to determine limits on the masses of planets interior to the cold belt. We determine that a minimum of four planets are required, each with a minimum mass of 0.64~M$_J$, in order to clear the gap between the asteroid and Kuiper belts of the system. If we make the assumption that the outermost planet is responsible for the stirring of the disc, the location of its inner edge and the eccentricity of the disc, then we can more tightly predict its eccentricity, mass and semimajor axis to be $e_{\rm{p}}=0.34^{+0.20}_{-0.13}$, $m_{\rm{p}}=0.7^{+0.5}_{-0.3}\,\rm{M}_{\rm{J}}$ and $a_{\rm{p}}=76^{+12}_{-13}\,\rm{au}$.
\end{abstract}


\section{Introduction}
As one of the key components of a planetary system, studying debris discs enables us to understand the current make up of a planetary system and its formation and evolution. In recent years the Atacama Large Millimeter/submillimeter Array (ALMA) has made it possible to image these discs at long wavelengths in much finer detail than was previously possible \citep[for a recent review see ][]{hughes18}. Of particular interest are systems where both debris discs and at least one planet have been observed. Analysis of such systems is of prime importance for understanding the interaction between planets and the disc. Examples of such systems include Fomalhaut \citep{kalas08,boley12}, HR~8799 \citep{marois10,booth16}, $\beta$ Pic \citep{lagrange10,dent14} and HD 95086 \citep{rameau13,su17a}. For systems where no planet has yet been directly imaged, strong constraints on where the outer planets in the system are can still be derived from studying the debris disc \citep[e.g. ][]{booth17,marino18,marino19}.

In this paper we present and analyse the first ALMA image of the debris disc around HD~38206. HD~38206 is a star of spectral type A0V. The debris disc around this star was first identified using IRAS data by \citet{mannings98} and has also been detected by \textit{Spitzer/MIPS} \citep{rieke05}, \textit{Spitzer/IRS} \citep{morales09}, Gemini/T-ReCS \citep{moerchen10} and \textit{Herschel/PACS} \citep{morales16}. By analysing both the resolved \textit{Herschel} images and the full SED, \citet{morales16} demonstrate that the system is seen close to edge on and has two belts at 11 and 160~au. The system is thought to be a member of the Columba association \citep{torres08}, giving it an age of 42$^{+6}_{-4}$~Myr \citep{bell15}. The young age of this system means that it is a prime candidate for direct imaging surveys, although no planets have been detected so far. \citet{shannon16} developed a model for the minimum mass of planets required to clear a gap in a two belt debris disc system. They use HD~38206 as an example case and show that the minimum mass of planets is close to the upper limit possible from VLT/SPHERE observations. By analysing the ALMA data of this system we shall re-assess the limits on the masses of potential planets in the system and make a prediction for the properties of the hypothetical outermost planet.

\section{ALMA Observations}
\label{sobs}
The observation of HD~38206 was carried out by ALMA in band~6 during cycle 1 as part of the project 2012.1.00437.S (PI: David Rodriguez). It was observed on the 7th\,March\,2014 with a precipitable water vapour of $1.80${~mm} and $35.7$~{mins} on source time by 23 antennas in a compact configuration. These led to baselines between $15-365$~{m}. The received data was calibrated using the standard observatory calibration in CASA version 4.7.74. The quasar J0609-1542 was used for bandpass and phase calibration and the active galactic nucleus PKS~0521-36 as flux calibrator. The correlator was set to provide four spectral windows, processing two polarizations in each of these. While one spectral window was centred at the CO $J$=2-1 line at $230.538$~{GHz}, with 3840~channels of width $0.5$~{MHz}, the other three had central frequencies of 213, 215 and 228 GHz and 128~channels with a width of $16$~{MHz} to study the dust continuum emission. 

\begin{figure}
    \centering
    \includegraphics[width=\columnwidth]{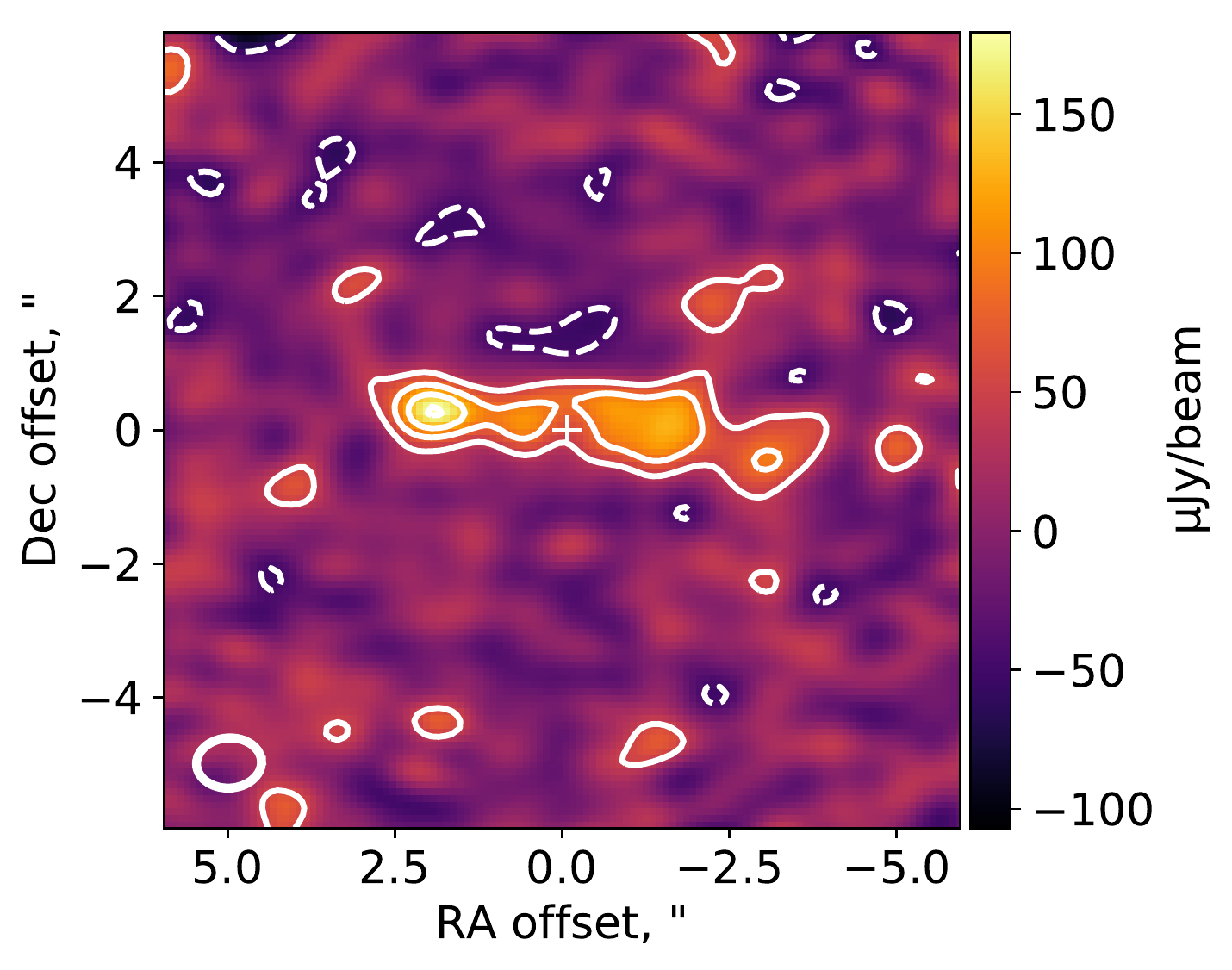}
    \caption{Image of HD~38206 at 1.35~mm after processing with the CLEAN algorithm. The contours show the $\pm 2,\, $4, 6 and 8$\sigma$ levels. The white cross marks the position on the star based on the \emph{Gaia} DR2 position and accounting for proper motion.}
    \label{fig:hd_dirtyimagebeam}
\end{figure}

The image of the disc is shown in \autoref{fig:hd_dirtyimagebeam}. This has been created from the inversely Fourier transformed complex visibilities using natural weighting and multi-frequency synthesis, followed by processing with the CLEAN algorithm \citep{hogbom74}. We obtain a synthesised beam of size $0.97{''}\times 0.75{''}$ and a beam position angle of $-86.5{\degr}$ measured from North to East. We measure the RMS to be $\sigma = 23.5$~{\ujypbm}. The disc is seen to be edge-on. There are some signs of asymmetry with a peak in the emission of 0.18$\pm$0.02~mJy 2.1$''$ east of the star, whilst to the west the emission peaks at 2.9$''$ from the star, but with a lower flux density of 0.09$\pm$0.02~mJy. We do not expect this to be due to a pointing issue as the phasecentre location agrees well with the expected position of the star based on the \emph{Gaia} DR2 \citep{gaia18} position after correcting for proper motion. 

The total flux density within an ellipse surrounding the emission is 0.7$\pm$0.1~mJy. \citet{morales16} found the flux density at 160~$\micron$ to be $188.9^{+6.5}_{-6.5}$~mJy. By assuming a single power law between 160~$\micron$ and 1350~$\micron$, typically formulated as $F_\nu=\lambda^{-(2+\beta)}$, we find $\beta=0.6$, a typical value for a debris disc \citep{holland17}.

To check for CO $J$=2-1 emission, we also used CLEAN to create a data cube of the channels around the expected radial velocity of the star \citep[25.3~km\,s$^{-1}$][]{gontcharov06}. No CO line emission was detected. By integrating over the pixels where continuum emission is detected (at $>3\sigma$) we find the $3\sigma$ upper limit on an unresolved emission line to be $1.4\times10^{-22}$~W\,m$^{-2}$ \citep[using Equation A2 of][]{booth19}. Based on a model of gas production through the collisional cascade, \citet{kral17} predicted that this system would have a CO $J$=2-1 level of $2.2\times10^{-25}$~W\,m$^{-2}$. Our upper limit is, therefore, consistent with this model and much deeper observations will be necessary to detect any gas in this system.

\section{Modelling}
\label{smod}
\begin{table}
	\caption[Stellar parameters of HD~38206.]{Overview of the stellar parameters that have been used for the analysis. }
	\label{tab:stellar_params}
	\centering
	{
	\begin{tabular}{lcc}
		Parameter & Value & Reference \\
		\hline
		$ d $, pc & 71.3$\pm$0.4 & 1 \\
		RA (J2000) & $  05^{\rm{h}}\, 43^{\rm{m}}\, 21.67^{\rm{s}} $ & 2 \\
		DEC (J2000)  & $ -18\degr\, 33'\, 26.91''  $ & 2 \\
		Age, Myr & $ 42^{+6}_{-4} $ & 3\\ 
		$ R_{\star} $, R$ _{\odot} $ & 1.7$\pm$0.2 & 4 \\
		$ T_{\textnormal{eff}} $, K & $ 9610^{+160}_{-1740} $ & 2 \\ 
		$ L_{\star} $, L$ _{\odot} $ & $ 26\pm7 $ & 4\\
		$ M_{\star} $, M$ _{\odot} $ & $ 2.4\pm0.4 $ & 4\\
		log($g$) & 4.4$\pm$0.3 & 4 \\
		
	\end{tabular}}
	{\small\begin{flushleft}
	\textbf{References.} (1)~\citet{bailer18}, (2)~\citet{gaia18}, (3)~\citet{bell15}, (4)~\citet{stassun18}
	\end{flushleft} \par}
\end{table}

\subsection{Disc setup} \label{sec:models}
The modelling procedure used here follows that of \citet{marino19}. We model the dust radial and vertical distribution in cylindrical coordinates, $\Sigma(r,\phi,z)$, as a Gaussian. Given that the image shows some signs of asymmetry, we define the radial distribution as a Gaussian in terms of the semi-major axis, $a(r,\phi)$, rather than the radial distribution, $r$, where the semi-major axis is determined by
\begin{equation}
 a(r,\phi) = r\frac{1-e\cos(\phi-\omega)}{1-e^2},
\end{equation}
where $e$ is the eccentricity and $\omega$ is the argument of pericentre. The surface density distribution is then given by
\begin{equation}
\Sigma(r,\phi,z) \propto \exp \left( -\frac{(a-a_0)^2}{2 \sigma_\textup{r}^2} -\frac{z^2}{2h^2r^2} \right)~, \\
\label{esdist}
\end{equation}
\begin{equation}
\sigma_\textup{r} = \frac{\Delta a}{2 \sqrt{2 \log 2}},
\end{equation}
where $ a_0 $ is the mean of the Gaussian, $ \Delta a $ is the full width half maximum (FWHM) of the Gaussian and $h$ is the aspect ratio. Given the low resolution of these observations, we do not expect to vertically resolve the disc and so have arbitrarily set \mbox{$h=0.01$}. 

\begin{table*}%
\caption{Parameters for the best fitting model. The uncertainties given for the parameters are the 16th and 84th percentiles of the posterior distribution. $\Omega$ is measured anti-clockwise from North, whilst $\omega$ is measured anti-clockwise from $\Omega$.}
\begin{tabular}{ccccccc}
{$ a_0 $, au} & {$ \Delta a $, au} &  {$ M_{\rm{dust}} $, M$_{\oplus}$} & {$I$, \degr} & {$\Omega$, \degr} & $e$ & $\omega$, \degr  \\
\hline
$184^{+19}_{-17}$ & $143^{+46}_{-36}$ & $0.105^{+0.016}_{-0.015}$ & $83.3^{+1.3}_{-1.3}$ & $84.3^{+1.2}_{-1.2}$ & $0.25^{+0.10}_{-0.09}$ & $49^{+22}_{-25}$  \\
\end{tabular}
\label{tab:hd_vis_params}
\end{table*}

\begin{figure*}
    \hspace{-1cm}\includegraphics[width=0.35\textwidth]{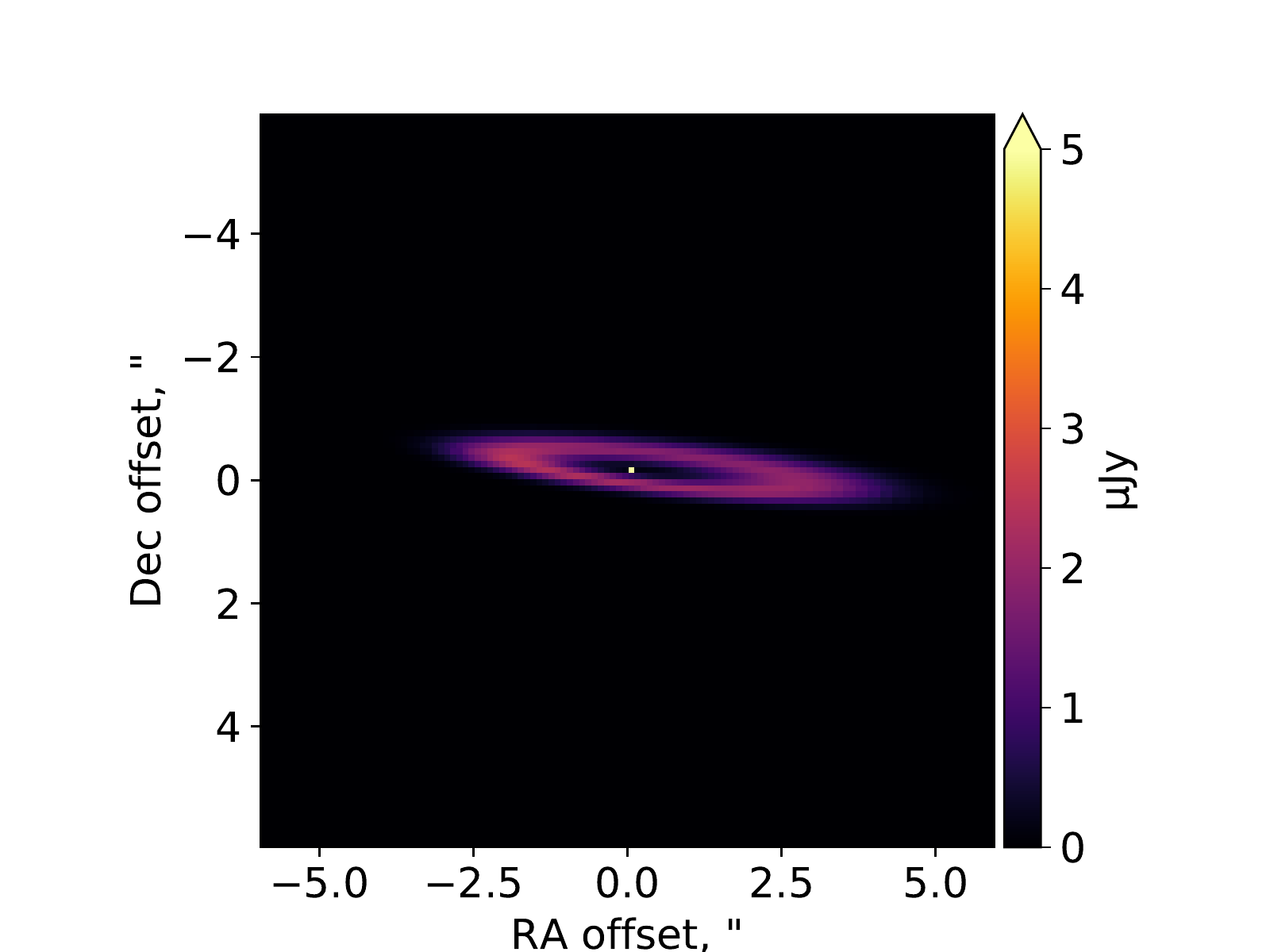}
    \hspace{-1cm}\includegraphics[width=0.35\textwidth]{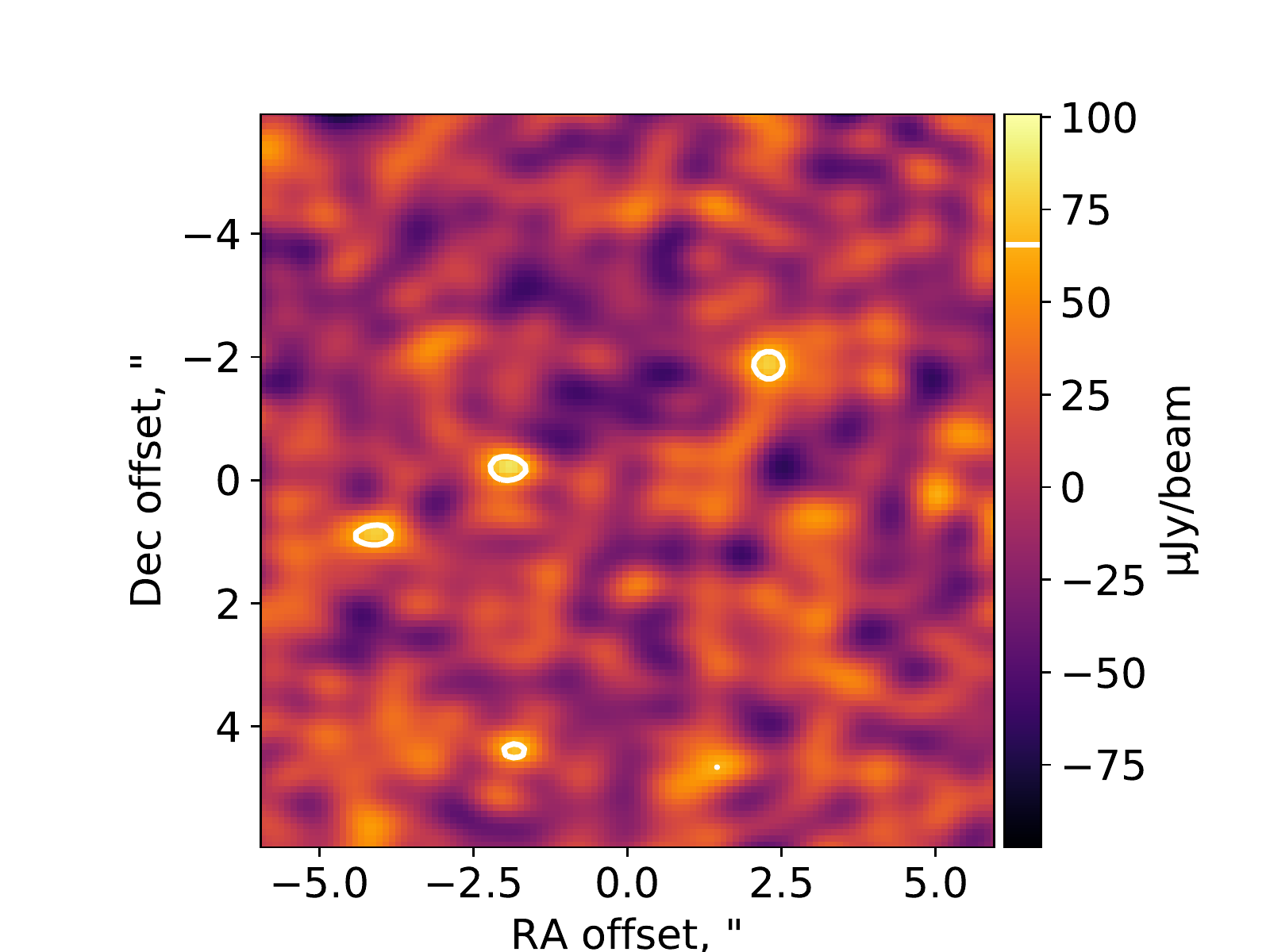}
    \includegraphics[width=0.35\textwidth]{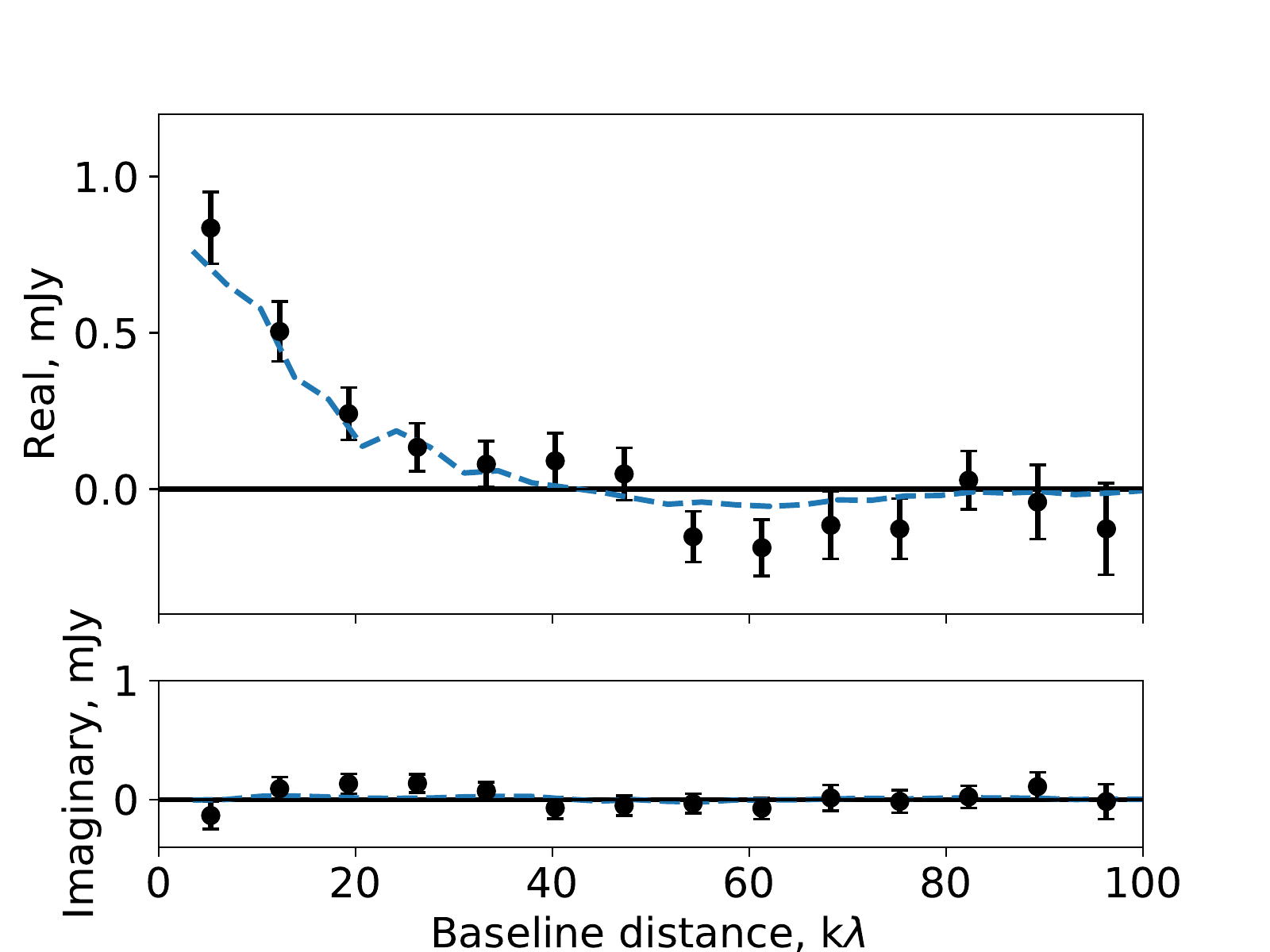}    
    \caption{\emph{Left: }Model image for the best-fitting parameters of the model. \emph{Middle: }Image residuals after subtracting the best-fitting model. The contours show the 3$\sigma$ residuals. \emph{Right: }The deprojected and azimuthally averaged observed visibilities (points) and model visibilities (dashed line).}
    \label{fig:hd_vis_plot}
\end{figure*}

For the dust grains we assume a differential size distribution $ n \propto s^{\alpha} $, with the power-law index $ \alpha~=~-3.5 $, following \citet{dohnanyi69}. Following the spectral energy distribution analysis of \citet{morales16}, we set the minimum grain radius to $S_{\rm{min}}=5\,\mu$m. The maximum grain radius is set to an arbitrarily large value of $S_{\rm{dust}}=1$\,cm. The total mass of dust in the disc is defined by $M_{\rm{dust}}$. For the optical properties, astrosilicate grains were assumed \citep{draine03} as these produced the best fit to the photometry \citep{morales16}. We note that, since we are modelling a resolved image at a single wavelength, these parameters have little effect on the final results, with the exception of the dust mass that defines the total flux density.

The stellar parameters are given in \autoref{tab:stellar_params}. To calculate the stellar flux we have used a Kurucz stellar template spectrum \citep{kurucz79} with values closest to those for HD~38206 ($T_{\textnormal{eff}} $=9500~K and log($g$)=4.5). 

To create the final image of the disc we define the inclination with respect to face-on, $I$, and the position angle anti-clockwise from North, $\Omega$, before running a radiative transfer computation using \rad \citep{dullemond12}.

To summarise, our model has seven free parameters -- $ a_0 $, $ \Delta a $, $M_{\rm{dust}}$, $I$, $\Omega$, $e$ and $\omega$.

\subsection{MCMC fit}

To find the best fitting parameters we make use of the \textit{emcee} module \citep{foreman13}. It is a Python implementation of an Affine Invariant Markov chain Monte Carlo (MCMC) Ensemble Sampler \citep{goodman10}. In this paper we fit directly to the visibility data, which has been averaged to one channel per spectral window. This means that we are fitting images at four different wavelengths, which is done since the differences between the spectral windows are large enough that simply averaging to one frequency could distort the final visibilities. However, this means that the model must be compared at four different wavelengths. Rather than create four separate models, one is created and adjusted to each wavelength assuming a power law change in the flux density with a $\beta=0.6$ for the disc and $\beta=0$ for the star (see \autoref{sobs}). The resultant image output from \rad{} is then multiplied by the primary beam, Fourier transformed and then sampled at the same points in the Fourier plane as the original data in order to compare the two. The $\chi^2$ statistic is then calculated using:
\begin{align} \label{echi2}
\chi^2 =& \sum_{j=1}^{N_{\textnormal{vis}}} 
\left(\mathfrak{Re}\left( D_j \right) 
- \mathfrak{Re}\left( M_j \right) \right)^2 w_j \nonumber \\
&+ \left(\mathfrak{Im}\left( D_j \right) 
- \mathfrak{Im} \left( M_j \right) \right)^2 w_j ~.
\end{align} 
where $D_j$ and $M_j$ are the observed and model visibilities respectively and $w_j$ are the weights.

A known problem with data taken during the early cycles of ALMA, such as the data analysed here, is that the weights are correct relative to each other but not necessarily correct in an absolute sense\footnote{\url{https://casaguides.nrao.edu/index.php/DataWeightsAndCombination}}. Using the uncorrected weights will still result in determining the best fit model but the uncertainties on the fitted parameters may be incorrect. To account for this we need to determine the factor, $f$, that the uncertainties are underestimated by. This could be left as a free parameter, however, since the SNR per visibility is much less than 1, we can make a very good estimate of $f$ by calculating the $\chi^2$ for a null model. Through this method $f=\sqrt{\chi^2/(2N_{\rm{vis}})}$, where $N_\textnormal{vis}$ is the number of visibilities used in the fit and we are using $2N_\textnormal{vis}$ because the real and imaginary components are independent \citep[see e.g. Section 6.2.2 of][]{thompson17}. The weights are then adjusted so that $ w_j \rightarrow \dfrac{w_j}{f^2} $.

Assuming Gaussian uncertainties, the likelihood is then:
\begin{equation} \label{elike}
\ln \mathcal{L}\propto-\chi^2/2.
\end{equation}
We used uniform priors for all free parameters with limits of $20$~au $ < r_0 < 350$~au, $10$~au $ < {\Delta}r  < 2r_0$, $60\degr  < I    < 90\degr$, $50\degr  < \Omega     < 130\degr$, $0.0001$~M$_{\oplus} < M_{\rm{dust}} < 0.3$~M$_{\oplus}$, $0<e<0.9$ and $-180\degr<\omega<180\degr$. The MCMC was run with 200 walkers and for 1000 steps.

\begin{figure}
    \centering
    \includegraphics[width=0.48\textwidth]{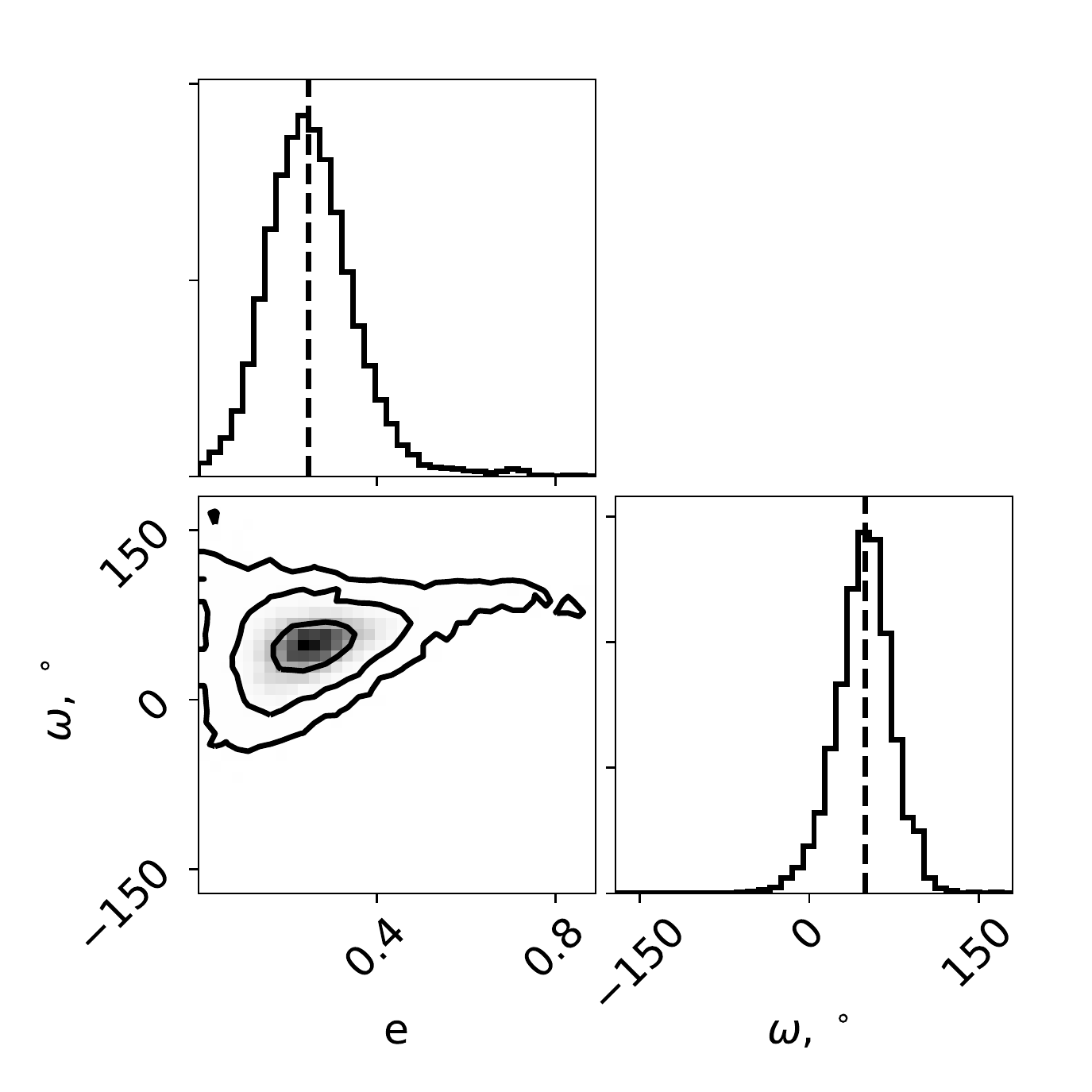}
    \caption{Posterior distribution for the eccentricity and the argument of pericentre. Whilst a non-zero eccentricity is clearly preferred, a circular ring is still consistent within the 3$\sigma$ uncertainties (defined by the outermost contour).}
    \label{fig:eocorner}
\end{figure}

\subsection{Results}
\label{sres}
The resulting best fitting parameters are given in \autoref{tab:hd_vis_params} whilst the best fitting model, residuals and deprojected visibilities are shown in \autoref{fig:hd_vis_plot}. The clearest difference between our results and the analysis of the Herschel data by \citet{morales16} is that we find the disc to be very wide whereas they assumed a narrow ring (of $\sim$20~au) and found it to be consistent with their data, which had a resolution of $\sim6\arcsec$. Nonetheless, our finding that the peak in the emission is at $184^{+19}_{-17}$~au is consistent with their narrow ring location. Our higher resolution data also clearly shows that this disc is close to edge on. \citet{morales16} found a lower inclination of $60\pm4\degr$, but this may also be a result of their use of a narrow ring model.

We find that a non-zero eccentricity is preferred by our fit, with a value of $0.25^{+0.10}_{-0.09}$. However, we cannot completely rule out a circular ring since that is still within the 3$\sigma$ uncertainty region (see \autoref{fig:eocorner}). The apocentre location on the west side of the disc allows this model to explain the extension to the west noted in \autoref{sobs}. It does not, however, explain the brightness asymmetry as seen by the 3$\sigma$ residual in the residuals image (\autoref{fig:hd_vis_plot}). This is not surprising as we do not expect to see any pericentre glow at these wavelengths \citep[e.g.][]{pan16}. Since the residuals image also shows a number of other 3$\sigma$ sources away from the disc, it is likely that these are simply peaks in the noise, although it is also possible that they are background sources.

Our model uses the dust mass as a free parameter. This can easily be converted to a flux density by summing the flux in the best fit image. Through this process we find that $F_{1350\,\mu\rm{m}}=$0.81$\pm0.12$~mJy, which is consistent with that found from the CLEAN image in \autoref{sobs}.

\section{Discussion}
When analysing the spatial distribution of a debris disc, the natural question to ask is what does this mean for the planetary system? 
From our and prior observations we know that: 
\begin{itemize}
 \item the very fact that we see dust at all requires that the planetesimals must have been stirred up enough to initiate a collisional cascade
 \item the radial distribution of the outer belt and the presence of an inner belt implies gap clearing in between the two
 \item there are tentative signs of asymmetry that imply dynamical interactions
\end{itemize}
In the following subsections we shall discuss whether or not these features necessitate the presence of planets in the system and make predictions for the properties of  of these potential planets. 

\subsection{Self-stirring}
\label{sstir}
In order to create a debris disc, the planetesimals remaining in a system after the gas of the protoplanetary disc has dissipated must be stirred so that the planetesimals can reach the relative velocities necessary to initiate a collisional cascade. The most likely ways for this to happen are via the gravitational effects of nearby planets \citep[planet stirring, e.g.][]{mustill09} or large planetesimals within the disc \citep[self-stirring, e.g.][]{kenyon08,krivov18b}. First we consider whether the disc around HD~38206 can be self-stirred during the age of the system. If large planetesimals form quickly in the disc (e.g. by streaming instability), \citet{krivov18b} show that
\begin{align}
T_{\rm{s}} &={9.3\,\rm{Myr}} \nonumber \\
  &\times\left( \frac{1}{\gamma} \right)  \left( \frac{\rho}{1\,\rm{g\,cm}^{-3}} \right)^{-1}  \left( \frac{v_\text{frag}}{30\,\rm{m\,s}^{-1}} \right)^4 \left( \frac{S_\text{max}}{200\,\rm{km}} \right)^{-3} \nonumber \\
  &\times\left( \frac{M_{\star}}{M_{\odot}} \right)^{-1/2}  \left( \frac{a_0}{100\,\rm{au}} \right)^{7/2}\left(\frac{\Delta a/a_0}{0.1}\right)\left(\frac{M_{\rm{disc}}}{100\,\rm{M}_\oplus}\right)^{-1}
\label{etstir}
\end{align} 
where $\gamma$ is a constant factor between 1 and 2, $\rho$ is the density, $v_\text{frag}$ is the velocity required for destructive collisions, $S_\text{max}$ is the maximum radius of planetesimals in the collisional cascade and $M_{\rm{disc}}$ is the total mass in the disc. 
For the composition we are using, $\rho = 3.5\,\rm{g\,cm}^{-3}$. From \autoref{sres} we find that $a_0=184$~au and $\Delta a=143$~au. 
We can estimate $M_{\rm{disc}}$ by extrapolating the dust mass assuming a single power law size distribution from dust grains up to $S_\text{max}$
\citep[although note that estimating the total disc mass is highly uncertain due a lack of knowledge in the form of the size distribution and its maximum extent; see][for further discussion on the total mass of debris discs]{krivov20}
\begin{equation}
 M_{\rm{disc}}=\frac{S_{\rm{max}}^{4+\alpha}-S_{\rm{min}}^{4+\alpha}}{S_{\rm{dust}}^{4+\alpha}-S_{\rm{min}}^{4+\alpha}}M_{\rm{dust}}.
\end{equation}
This results in $M_{\rm{disc}}=420\,\rm{M}_\oplus$. For all other parameters we use the standard values used by \citet{krivov18b}: $\gamma=1.5$, $v_\text{frag} = 30\,\rm{m\,s}^{-1}$ and $S_\text{max} = 200\,\rm{km}$. This results in $T_{\rm{stir}} = 13\,\rm{Myr}$. 

HD~38206 has previously been reported as a member of the Columba association \citep{torres08}. We first check that this is still the case using the latest \emph{Gaia} data and the bayesian inference association membership code BANYAN $\Sigma$ \citep{gagne18}. This results in a 99.9\% chance that this star is a member of the Columba association. According to \citet{bell15}, this means that it has a likely age of 42$^{+6}_{-4}$~Myr, longer than the timescale needed for self-stirring and demonstrating that planets are not required to stir the disc.

However, we must note that there is a lot of uncertainty in many of these values. For instance, even a small change in the slope of the size distribution can result in a large change in total disc mass.
The timescale for stirring is even more strongly dependent on $v_\text{frag}$ and $S_\text{max}$. Laboratory experiments demonstrate that there is an uncertainty of about an order of magnitude on $v_\text{frag}$ \citep[see][and references therein]{blum08}, whilst $S_\text{max}$ is dependent on the assumption that the planetesimals formed through the process of pebble concentration described in \citet{johansen15,simon16,schafer17,simon17}. Bearing all  this in mind, there is at least a couple of orders of magnitude uncertainty on the self-stirring timescale and so we should also consider the possibility that planets are required to stir the disc.

\subsection{Gap clearing by planets}
\label{sgap}
Like many other debris discs, the one around HD~38206 has a two-component
structure -- in addition to the belt on the outer edge of this system imaged by ALMA, we know from the SED \citep{morales09, morales16} that there is also warm emission likely originating from an asteroid belt analogue. Both the origin of the broad gap between the outer and the inner rings
and the nature of the warm component in such disks are a matter of debate.
It is possible that planetesimals failed to form at intermediate distances, creating a
two-belt architecture by the time of gas dispersal
\citep[e.g.,][]{carrera17}, although this depends strongly on the planetesimal
formation model.
Another conceivable explanation for two-component discs
would be a swarm of planetesimals in eccentric orbits with apocentres in the outer belt
and pericentres in the inner one \citep{wyatt10},
although this implies high eccentricities that would be difficult to explain.

Nevertheless, by analogy with the Solar system with its Kuiper and asteroid belts and
giant planets
in between, it is natural to expect that the gap between the two rings was carved by as
yet undiscovered planets.
It would be interesting to put some constraints on the number, masses, and location
of the alleged planets in the cavity.
Such constraints would be tighter if the disc exhibited strong asymmetries
\citep[e.g.,][]{lee16,lohne17}, 
or if an accurate radial profile of the inner edge of the outer disc could be inferred from
the resolved images \citep[e.g.,][]{nesvold15,booth17}.
Also, for systems with the observed hot dust close to the star, constraints can be derived
by assuming this dust to stem from comets scattered into the inner system by a chain
of planets interior to the outer belt \citep[e.g.,][]{marino18b}.
However, none of this is the case for HD~38206,
so that we can only use the gap radius to constrain
parameters of the possible planets in this system.
This makes the problem highly degenerate,
since the same gap could be cleared by many different planet configurations.

In view of this degeneracy, we make a number of assumptions. All of them are arbitrary and
do not have to hold in reality, yet they allow us to make specific estimates.
Again by analogy with our Solar system, we confine ourselves to a simple scenario
in which the presumed planets in the gap are all in nearly-circular,
nearly-coplanar orbits \citep[e.g.,][]{su14},
although alternative scenarios exist,
such as
divergent planetesimal-driven migration of pairs of planets
\citep{morrison18} or
sweeping secular resonances with one `lonely' giant planet in an eccentric orbit
\citep{zheng17}.
For simplicity, we also assume a log-uniform spacing and equal masses for all
planets in the gap.
Finally, we assume that the alleged planets are dynamically stable against mutual
perturbation over the age of the central star.

\begin{figure}
	\includegraphics[width=\columnwidth]{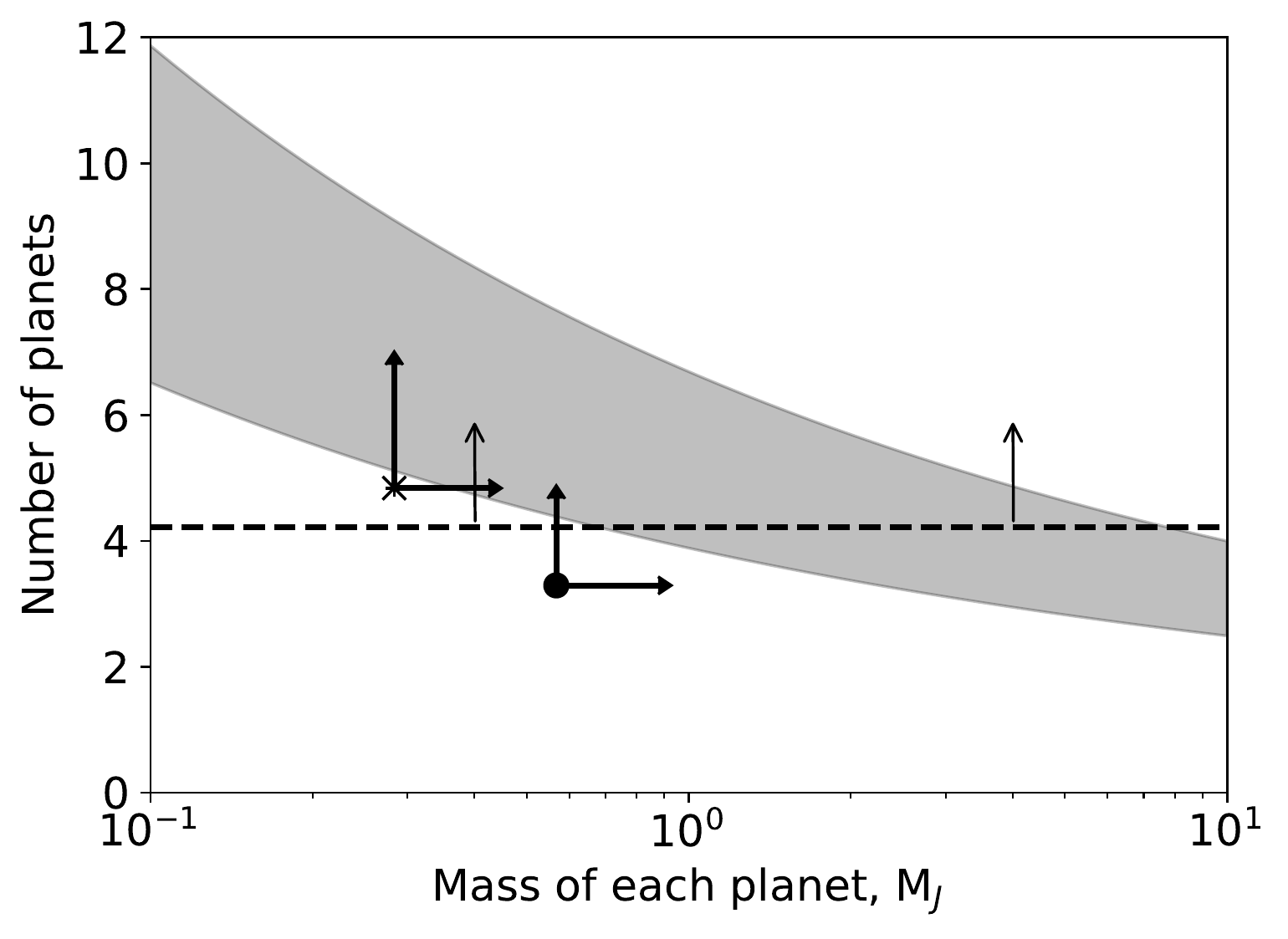}
    \caption{Constraints on the possible number of planets in the gap and their masses. The grey shaded region represents the bounds based on the model of \citet{faber07} (see Equations \ref{efq1} and \ref{efq2}). The cross and the dot represent the minimum number and mass based on the model of \citet{shannon16} and using $K=16$ and $K=20$ respectively (see Equations \ref{eshannonmass} and \ref{eshannonno}). The horizontal dashed line represents the minimum number below which the equations of \citet{shannon16} are invalid (see Equation \ref{econdition}).}
    \label{fig:planet_constraints}
\end{figure}

Many studies derived stability criteria for both lower mass planets
\citep[e.g.,][among others]{chambers96,faber07,%
zhou07,smith09b} and massive ones
\citep{morrison16}.
One way of applying these criteria to the gap clearing was proposed by
\citet{faber07}. First they estimate the \emph{maximum} number of planets $N_\mathrm{p}$
in relation to their mass $m_\mathrm{p}$ that would still lead to a (marginally) stable
system. Since the planets themselves would already be close to becoming unstable, they
argue that planetesimals would already be completely removed from the planetary region,
ensuring the gap to be devoid of any debris material. 
They then also estimate the \emph{minimum} number of planets. To do this,
they place them at separations that are twice as large as
those that correspond to a marginally stable system. In this case, the expectation is that
a significant fraction of planetesimals residing between the orbits
of neighbouring planets would be removed from the system. It is only the bodies
orbiting exactly half-way between the planets that would have a chance to survive.
The planetary system would then certainly be stable,
while the gap would be nearly free of debris. 
In order to calculate these extremes, \citet{faber07} provide the following formulae:
\begin{equation}
 N_p=\frac{\log_{10}(a_{\rm{in}}/a_{\rm{w}})}{\log_{10}(1+g\delta_{\rm{min}})}
 \label{efq1}
\end{equation}
\begin{equation}
 \delta_{\rm{min}}=\frac{\mu^{1/4}}{3.7}(\log_{10}(\tau/\rm{yr})+1+\log_{10}(\mu/10^{-7}))
 \label{efq2}
\end{equation}
where $a_{\rm{in}}$ is the inner edge of the outer belt, $a_{\rm{w}}$ is the location of the inner belt (which is assumed to be narrow), $g$ is a parameter that equals 1 for calculating the maximum number of planets and 2 for calculating the minimum number of planets, $\mu$ is the mass ratio $m_p/M_\star$ and $\tau$ is the age of the system. The relationship between $N_p$ and $m_p$ using the parameters for the HD~38206 system is shown in \autoref{fig:planet_constraints} with the grey region representing the allowed parameters. 

Based on these results, we might expect between 4 and 7 Jupiter-mass planets. A larger number of lower mass planets
would also do. Note that there is no stringent lower limit on
the mass of a single planet in the model, except that the underlying stability criterion
was only established by \citet{faber07} by numerical integrations within a certain
range of masses and may fail outside that range.

Another approach is to put the suspected planets in a configuration that is
spaced widely enough to ensure stability, place planetesimals between the planetary
orbits, and to invoke numerical integrations and analytic arguments to see
whether, and on which timescales, these planetesimals will be removed from the system.
This idea was employed by \citet{shannon16} who found
the clearing timescale of the chaotic zone as a function of the planet mass
and semimajor axis using $N$-body simulations. From these simulations they derived a lower limit on the mass of each planet, $m_{\rm{p}}$, inside the gap, assuming equal mass planets with a separation of $K$ mutual Hill radii. Modifying their equations such that $K$ is left as a free parameter, we find that:
\begin{equation}
m_{\rm{p}} = \left( \dfrac{\kappa}{\tau} \right) \left( \dfrac{a_{\rm{in}}}{1\,\rm{{au}}} \right)^{3/2} \left( \dfrac{M_{\star}}{M_{\odot}} \right)^{1/2} m_{\oplus}~,
\label{eshannonmass}
\end{equation}
and also the minimum number of planets that are needed to clear the gap within the age of the system:
\begin{equation} 
N_{\rm{p}} = 1 + \dfrac{ \log_{10} \left( \frac{a_{\rm{in}}}{a_{\rm{w}}} \right) }{ \log_{10}  \left( \dfrac{1+0.006K  \left( \frac{m_{\rm{p}}}{m_{\oplus}} \right)^{1/3} \left( \frac{M_{\star}}{M_{\odot}} \right)^{-1/3}}{ 1-0.006K  \left( \frac{m_{\rm{p}}}{m_{\oplus}} \right)^{1/3} \left( \frac{M_{\star}}{M_{\odot}} \right)^{-1/3} } \right) }~,
\label{eshannonno}
\end{equation}
where $\kappa$ is a parameter dependent on $K$.
They first assumed a wide spacing, with separation of $K = 20$
mutual Hill radii between the neighbouring planet orbits, which is close to a typical
separation in Kepler multiplanet systems. In this case $\kappa=(4\pm1)\times10^{6}$~yrs. They also checked a slightly tighter packing,
with $K=16$, and found the results to be similar. In this case $\kappa=(2\pm0.2)\times10^{6}$~yrs. 

\citet{shannon16} applied their formulae to the HD~38206 system using the parameters known at the time ($\tau=30$~Myr, $a_{\rm{w}}=15$~au, $a_{\rm{in}}=180$~au). They found that 3 planets, each with a mass of 1.4~M$_J$, are required to clear the gap. 

In their case, $a_{\rm{in}}$ was determined solely from a fit to the SED and assumption of a narrow belt. From our results, whilst we do not resolve the inner edge of the disc, we do resolve the disc well enough to determine that it must be much wider than previously assumed. For our purposes, we shall approximate the edges of the disc with the FWHM of the fitted Gaussian. In other words, $a_{\rm{in}}=a_0-\Delta a/2=110\pm30$~au. Whilst follow-up, higher resolution observations may determine that the real distribution is different to this, the uncertainties on the inner edge measured here are still quite large and so likely to contain the real inner edge\footnote{For example, if the real distribution was closer to a boxcar function with sharp edges, then it can be demonstrated that fitting a Gaussian function to this results in an inner edge located at $a_0-(0.85\Delta a)/2=123$~au. Therefore, in this example the estimates of the mass and number of planets would be closer to the upper limit of the predictions we have made.}. In addition we also update the other two parameters, now using $\tau=42^{+6}_{-4}$~Myr \citep{bell15} and $a_{\rm{w}}=8.7$~au \citep{morales16} 
From this we find (assuming Gaussian uncertainty propagation)
$N_\mathrm{p}=3.3\pm1.4$ and $m_\mathrm{p}=0.6\pm0.3 M_\mathrm{J}$ for $K=20$ and
$N_\mathrm{p}=4.8\pm2.0$ and $m_\mathrm{p}=0.28\pm0.14 M_\mathrm{J}$ for $K=16$
(large filled circle and cross, respectively,
in Fig.~\ref{fig:planet_constraints}). In other words, since our data allows us to determine that the disc is wide -- something that was not possible with the \emph{Herschel} data -- we find that the gap is smaller than the gap assumed by \citet{shannon16} and so the masses required to clear the gap are lower than given by their analysis.

However, these results should be treated with caution. \citet{shannon16} showed that at some point \autoref{eshannonmass} breaks down, and increasing the planet mass beyond this point no longer reduces the clearing time (their Figures 2 and 4). Systems are particularly susceptible to this effect if they are young, or have a distant outer belt. Specifically this happens when:\begin{equation}
    N_{\rm{p}}/2 - 1 > \log_{10}\left( \frac{a_{\rm{in}}}{a_{\rm{w}}} \right)~.
    \label{econdition}
\end{equation}
For our parameters we find that at least 4 planets are necessary to satisfy this condition. This criterion is, therefore, satisfied when we assume $K=16$, but not when we assume $K=20$. This implies that for a wide spacing of planets, it is not possible for any planetary system to clear the gap in such a short time, no matter how massive they are. Nonetheless, even in this case, there are reasonable uncertainties on several parameters in Equations \ref{eshannonmass} and \ref{eshannonno}, and planets with masses at the lower end of the uncertainty interval would satisfy \autoref{econdition} and could therefore clear the gap within the system age. In summary, if multiple equal-mass planets are to have cleared the gap, then this implies that either the location of the inner edge of the outer disc lies at the lower end of our calculated uncertainty interval, that the system age and/or stellar mass is at the upper end of the allowed range, or that the planets have a spacing that is smaller than 20 mutual Hill radii.

These estimates are very conservative compared to those obtained with the
\citet{faber07} method, providing a \emph{very} lower limit on the
total number of planets in the cavity.
The reason is a wide spacing of $K = 16$ or $20$ is
assumed by \citet{shannon16}.
For comparison, the \citet{faber07} curve for the maximum number of planets
corresponds to values of $K$ between 6 and 7, and the curve for the minimum number of planets to values of $K$ between 12 and 14.

\subsection{Constraints on the outermost planet}
\begin{figure}
    \centering
    \includegraphics[width=0.48\textwidth]{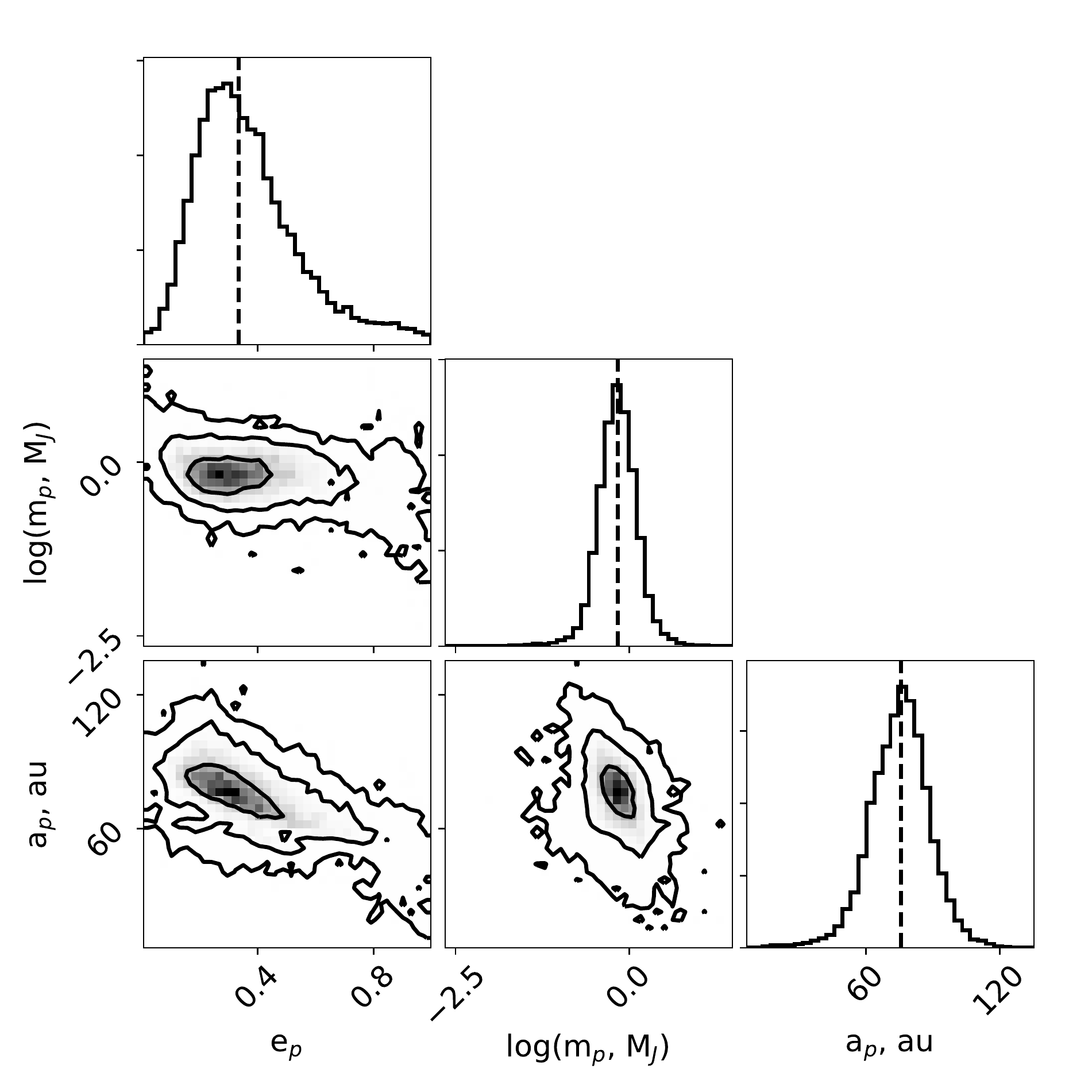}
    \caption{Posterior distribution for the eccentricity, mass and semi-major axis of the predicted outermost planet based on the assumption that it is responsible for stirring the disc, clearing material out to the inner edge and forcing the eccentricity of the disc particles. The three equations (\ref{epstir}, \ref{einner} and \ref{eplanete}) are then solved simultaneously for each of the sets of disc parameters given by the MCMC samples in \autoref{smod}.}
    \label{fig:plcorner}
\end{figure}

As noted in \autoref{sstir}, an alternative possibility to the disc being stirred by the largest planetesimals is that the disc is stirred by planets in the system. This is most likely to be due to secular perturbations from the outermost planet. The timescale for planet stirring is defined by \citet{mustill09} as
\begin{align}
 t_{p}&\approx1.53\times10^3\frac{(1-e_{\rm{p}}^2)^{3/2}}{e_{\rm{p}}}\left(\frac{a_{\rm{out}}}{10\,\rm{au}}\right)^{9/2} \nonumber \\
 &\times\left(\frac{M_\star}{\rm{M}_\odot}\right)^{1/2}\left(\frac{m_{\rm{p}}}{\rm{M}_\odot}\right)^{-1}\left(\frac{a_{\rm{p}}}{1\,\rm{au}}\right)^{-3} \rm{yr} 
 \label{epstir}
\end{align}
where $a_{\rm{out}}=a_0+\Delta a$, $e_{\rm{p}}$ is the eccentricity of the planet and $a_{\rm{p}}$ is the semimajor axis of the planet. Therefore, if we assume that this timescale is equivalent to the age of the system and the outermost planet in the system is solely responsible for stirring the disc out to its outer edge, we can find a family of solutions for its mass, semi-major axis and eccentricity. Strictly speaking, these are all lower limits since a more massive, more distant or more eccentric planet could stir the disc on a timescale shorter than the age of the system.

In addition, following \autoref{sgap}, we can consider the impact that the outermost planet will have on the inner edge of the disc. This is a problem that has been studied by multiple authors over the years \citep[see e.g.,][]{wisdom80,mustill12,pearce14,nesvold15}. Here we use the formulas of \citet{pearce14}, which take into account the eccentricity of the planet. Combining their equations 9 and 10, we find:
\begin{equation}
 m_{\rm{p}}=\frac{1}{125}\left(\frac{a_{\rm{in}}}{a_{\rm{p}}(1+e_{\rm{p}})}-1\right)^3(3-e_{\rm{p}})M_{\star}
 \label{einner}
\end{equation}
Thus we now have two equations, with three unknowns.

One of the unknowns is the planet eccentricity. If the planet is eccentric, then this will affect the disc eccentricity. Therefore our constraints on the disc eccentricity can provide constraints on the planet eccentricity. The two can be related through the equation \citep{mustill09}:
\begin{equation}
 e_{\rm{p}}=\frac{4}{5}\frac{a}{a_{\rm{p}}}e.
 \label{eplanete}
\end{equation}
In our modelling we have assumed a single eccentricity for the disc. In fact, as can be seen from this equation, the influence of a planet on a planetesimal's eccentricity is stronger the closer the planetesimal is to the planet. Therefore, for a wide disc, such as that of HD~38206, the disc will have a greater eccentricity at its inner edge than at its outer edge. However, the resolution of our observations is not good enough to detect such a variation in the eccentricity and so we assume a single eccentricity throughout and set $a=a_0$.

Therefore, if we make the assumption that the outermost planet is simultaneously responsible for the stirring of the disc, the clearing of the inner edge and the eccentricity of the disc, then we can solve Equations \ref{epstir}, \ref{einner} and \ref{eplanete}. Due to the large uncertainty in our knowledge of the disc parameters, we make use of all of the samples from our MCMC run,
solving the simultaneous equations for each set of disc parameters. Doing so we determine that such a planet would have $e_{\rm{p}}=0.34^{+0.20}_{-0.13}$, $m_{\rm{p}}=0.7^{+0.5}_{-0.3}\,\rm{M}_{\rm{J}}$ and $a_{\rm{p}}=76^{+12}_{-13}\,\rm{au}$. The full posterior distribution is shown in \autoref{fig:plcorner}.

An attempt to find planets in the system has been made by the NaCo Imaging Survey for Planets around Young stars (ISPY) using the Very Large Telescope \citep{launhardt20}. No planets were detected, nonetheless they find that the most stringent limits are for planets at distances of $\gtrsim$70~au, for which they rule out planets of masses $\gtrsim 7$~M$_J$ (Launhardt et al., in prep.). Interior to this the upper limits are less constraining. It is also important to note that, given that the disc is close to edge on, this only applies to planets that are close to the disc ansae. Assuming that any planets are coplanar with the disc, planets more massive than these limits could have evaded detection, particularly if they are on parts of their orbits that place them inside the inner working angle of the observations. An improvement in mass detection sensitivity of about an order of magnitude will be necessary (which requires an improvement of roughly 5 orders of magnitude in contrast) in order to be certain to detect the expected planets in the system and, given the high inclination, a low inner working angle will also be necessary to avoid occluding them when they are not near the disc ansae.

\section{Conclusions}
In this paper we present archival ALMA observations, taken at a wavelength of 1.3~mm, of the debris disc around the A0V star HD~38206. The data show the disc to be clearly resolved and highly inclined, close to edge-on. We also note some signs of asymmetry, with the disc extending further to the west than it does to the east. Using an MCMC analysis, we determine the dust to peak at a distance of 184$^{+19}_{-17}$~au with a width of 143$^{+46}_{-36}$~au and find that the asymmetry is best fit by an eccentricity of 0.25$^{+0.10}_{-0.09}$. 

The extreme width of the disc naturally leads to the question of whether the disc can be self-stirred or whether planets are necessary to initiate the collisional cascade. Using the equations of \citet{krivov18b} and their canonical parameters, we find that $\sim$16~Myr is necessary for self-stirring, which is less than the estimated age of the system, demonstrating that self-stirring should suffice to stir the disc out to its outer edge within the age of the system. Nonetheless, planets may well be required to explain the gap between the cold dust and the warm dust detected by \emph{Spitzer}. Based on the geometry of the system and the equations of \citet{shannon16}, we determine that a minimum of four planets are required, each with a mass of 0.28$\pm$0.14~M$_J$. 

By making the assumption that the outermost planet is responsible for the features of the disc -- its stirred nature, inner edge and eccentricity -- we predict that it would need to have $e_{\rm{p}}=0.34^{+0.20}_{-0.13}$, $m_{\rm{p}}=0.7^{+0.5}_{-0.3}\,\rm{M}_{\rm{J}}$ and $a_{\rm{p}}=76^{+12}_{-13}\,\rm{au}$. To detect such a planet requires an order of magnitude improvement in mass detection sensitivity over current observing capabilities.

\section*{Acknowledgements}
MB, AK and TP acknowledge support from the Deutsche Forschungsgemeinschaft through projects Kr 2164/13-2, Kr 2164/14-2 and Kr 2164/15-2. The authors thank the referee for their constructive report.

ALMA is a partnership of ESO (representing its member states), NSF (USA) and NINS (Japan), together with NRC (Canada), MOST and ASIAA (Taiwan), and KASI (Republic of Korea), in cooperation with the Republic of Chile. The Joint ALMA Observatory is operated by ESO, AUI/NRAO and NAOJ. In addition, publications from NA authors must include the standard NRAO acknowledgement: The National Radio Astronomy Observatory is a facility of the National Science Foundation operated under cooperative agreement by Associated Universities, Inc.
 
\section*{Data availability}
This paper makes use of the following ALMA data: ADS/JAO.ALMA\#2012.1.00437.S. Data resulting from this work are available upon reasonable request.

\bibliographystyle{mnras}
\bibliography{thesis}{}



\bsp

\end{document}